\documentstyle[psfig]{mn}

\title{Dynamical state of superclusters of galaxies: do
superclusters expand or have they started to collapse?}
\author[Mirt Gramann and Ivan Suhhonenko]
{Mirt Gramann and Ivan Suhhonenko
   \\
   Tartu Observatory,
       T\~oravere 61602, Estonia}
\begin{document}
\maketitle

\let\sec=\section
\let\ssec=\subsection
\let\sssec=\subsubsection

\def\kms{\;{\rm km\,s^{-1}}}
\def\kmsmpc{\;{\rm km\,s^{-1}\,Mpc^{-1}}}
\def\hompc{\,h\,{\rm Mpc}^{-1}}
\def\mpcoh{\,h^{-1}\,{\rm Mpc}}
\def\mpc3h{\,h^{3}\,{\rm Mpc^{-3}}}

\begin{abstract}
We investigate the dynamical state of superclusters in Lambda
cold dark matter ($\Lambda$CDM) cosmological models, where the
density parameter $\Omega_0=0.2-0.4$ and $\sigma_8$ (the rms
fluctuation on the $8h^{-1}$Mpc scale) is $0.7-0.9$. To study the
nonlinear regime, we use N-body simulations. We define
superclusters as maxima of the density field smoothed on the
scale $R=10h^{-1}$Mpc. Smaller superclusters defined by the density field
smoothed on the scale $R=5h^{-1}$Mpc are also investigated. We
find the relations between the radially averaged peculiar velocity
and the density contrast in the superclusters for different
cosmological models. These relations can be used to estimate the
dynamical state of a supercluster on the basis of its density
contrast. In the simulations studied, all the superclusters
defined with the $10h^{-1}$Mpc smoothing are expanding by the present
epoch. Only a small fraction of the superclusters defined with
$R=5h^{-1}$Mpc has already reached their turnaround radius and
these superclusters have started to collapse. In the model with
$\Omega_0=0.3$ and $\sigma_8=0.9$, the number density of objects
which have started to collapse is $5 \times 10^{-6}
h^3$Mpc$^{-3}$.

The results for superclusters in the N-body simulations are compared
with the spherical collapse model. We find that the radial peculiar
velocities in N-body simulations are systematically smaller than
those predicted by the spherical collapse model ($\sim 25$\% for
the $R=5h^{-1}$Mpc superclusters).

\end{abstract}


\sec{INTRODUCTION} Superclusters of galaxies are the largest
coherent and massive structures known in the Universe. The
existence of superclusters is known since the pioneering studies of
Shapley (1930). The nearest example is the Local Supercluster
with the Virgo cluster as the central cluster (de Vaucouleurs
1956). The shape, mass and dynamical state of different
superclusters have been studied in several papers (see e.g.
Einasto et al. 1997; Ettori, Fabian \& White 1997; Small et al.
1998; Barmby \& Huchra 1998; Batuski et al. 1999; Bardelli et al.
2000; Basilakos, Plionis \& Rowan-Robinson 2001; and references
therein). Ettori, Fabian \& White (1997) studied the mass
distribution in the Shapley Supercluster, using X-ray
observations. To investigate the dynamical state of the observed
overdense regions they used the spherical collapse model (Gunn \&
Gott 1972; Peebles 1980). Ettori, Fabian \& White (1997) found
that the core region of the Shapley Supercluster with the radius
$6.7h^{-1}$Mpc is close to the turnaround point, when the
perturbed region ceases to expand and begins to collapse.
Bardelli et al. (2000) also analyzed the
dynamical state of the Shapley Supercluster, using the spherical
collapse model. They estimated that the total overdensity of
galaxies in the Shapley Supercluster is ($N/\bar N$) $\sim 11.3$
on a scale of $10.1h^{-1}$Mpc and concluded that, if light traces
mass and the density parameter $\Omega=1$, the Shapley
Supercluster has already reached its turnaround radius and has
started to collapse. Small et al. (1998) studied the structure
and dynamics of the Corona Borealis Supercluster. They found that
this supercluster may have started to collapse.

In this paper we study the dynamical state of superclusters in
Lambda cold dark matter ($\Lambda$CDM) cosmological models, using
N-body simulations. These models successfully explain many
observations of the large- and small-scale structure including
the mass function and the peculiar velocities of clusters of
galaxies (see e.g., Ostriker \& Steinhardt 1995; Gramann \&
H\"utsi 2001). We study the radial velocity and the density
contrast in superclusters. The spherically averaged radial
velocity around a system, in the shell of radius $R$, can be
written as
$$
u = HR - v,
\eqno(1)
$$
where $v_H=HR$ is the Hubble expansion velocity and $v$ is
the averaged radial peculiar velocity toward the centre of the system.
At the turnaround point, the peculiar velocity $v=HR$ and $u=0$.
If $v<HR$, the system expands and if $v>HR$, the system begins to
collapse. In the spherical collapse model, the peculiar velocity $v$
is directly related to the density contrast $\delta$ (see below).
We study the relation between $v$ and $\delta$ in N-body simulations
and compare the results with the spherical collapse model. Can we
estimate the dynamical state of the supercluster on the basis of
its density contrast?

The velocity field around superclusters and clusters in different
N-body simulations has been studied in several papers
(Lee, Hoffman \& Ftaclas 1986; Villumsen \& Davis 1986;
Lilje \& Lahav 1991; van Haarlem \& van de Weygaert 1993;
Hanski et al. 2001). Villumsen \& Davis (1986) studied the nature of
the velocity field around large clusters in $\Omega=1$ CDM models.
They found that the flowfields when averaged over $4\pi$ sr fit
spherical collapse model relatively well for density contrasts
$\delta<3$, but the mean radial peculiar velocity is systematically
low for increasing $\delta$. van Haarlem \& van de Weygaert (1994)
studied the evolution of the velocity profile around clusters
in $\Omega=1$ CDM models. They found that for different clusters
the agreement with the spherical collapse model can be different.
There are clusters where the agreement is very good, but often the
predictions of the spherical collapse model compare badly with the
actual velocity field in N-body simulations (see Fig. 16-17 in their
study). Hanski et al. (2001) studied the velocity profile around
four clusters in a $\Omega_0=0.3$ $\Lambda$CDM model. They found
that the radially averaged velocity fields around simulated clusters
are compatible with the spherical collapse model.

In this paper we study the velocity field for superclusters
in flat $\Lambda$CDM models with the density parameter
$\Omega_0=0.2-0.4$, the baryon density $\Omega_b h^2=0.02$ and the
normalized Hubble constant $h=0.7$. These values are in agreement with
measurements of the density parameter (e.g. Bahcall et al. 1999), with
measurements of the baryon density from abundances of light elements
(O'Meara et al. 2001; Tytler et al. 2000) and with measurements of the
Hubble constant using various distance indicators (Freedman et al.
2001; see also Parodi et al. 2000). To restore the spatial flatness in
the low-density models, we assume a contribution from the cosmological
constant $\Omega_{\Lambda}=1-\Omega_0$. We also assume that the initial
density fluctuation field is a Gaussian density field. In this case,
the power spectrum provides a complete statistical description of the
field.

We define superclusters as maxima of the density
field smoothed on the scale $R=10h^{-1}$Mpc. Smaller
superclusters in the density field smoothed on the scale
$R=5h^{-1}$Mpc are also investigated. We study the number density
of superclusters, where the radially averaged peculiar velocity is
larger than a given value, $n(>v)$. In particular, we investigate
the number density of superclusters which have radially averaged
peculiar velocities $v>HR$.

This paper is organized as follows. In Section~2 we consider the
spherical collapse model in a flat universe with cosmological
constant. We find the relation between the peculiar velocity and
the density contrast for the values of the density parameter
$\Omega_0=0.2-0.4$. In Section~3 we describe the N-body
simulations and the algorithms that have been used to identify
the superclusters. In Section~4 we investigate the density
contrast and the radial peculiar velocity for the superclusters
in the N-body simulations and compare the results with the
spherical collapse model. The function $n(>v)$ is analyzed in
Section~5 and a summary is presented in Section~6.

\sec{THE SPHERICAL COLLAPSE MODEL}
The spherical collapse model describes the evolution of a spherically
symmetric perturbation in an expanding universe. Under the assumption
of sphericity, the nonlinear dynamics of a collapsing shell is
determined by the mass interior to it. The spherical collapse model has
been discussed in detail by Tolman (1934), Bondi (1947), Gunn \& Gott
(1972), Silk (1974, 1977), Peebles (1980) and Schechter (1980).
This model has been used in the Press-Schechter (Press \& Schechter 1974)
formalism to evaluate the mass function of collapsed objects in the
universe, as well as to determine the density parameter of the universe
from the infalling flow (e.g. Davis \& Huchra 1982). In recent years,
the spherical collapse model has been used to study the dynamical state
of superclusters of galaxies (Ettori, Fabian \& White 1997;
Small et al. 1998; Bardelli et al. 2000).

Although the Zel'dovich approximation (Zel'dovich 1970) generically yields
pancakes and numerical simulations show that many collapsing regions are
filamentary (e.g. Gramann 1988; Bertchinger \& Gelb 1991), the spherical
collapse model has been popular because of its simplicity. Despite its
idealized nature, the spherical model appears to give a reasonable
description of collapse of high peaks in a random Gaussian density field
(Bernardeau 1994). Eisenstein \& Loeb (1995) studied an analytical model
for the triaxial collapse of cosmological perturbations. They found that
the turn-around times of the short axes of a triaxial perturbation are
fairly well predicted by the turn-around time of a spherical perturbation
with the same initial density.

We consider the spherical collapse model in a flat universe with
a cosmological constant, $\Lambda$. This model was studied by
Peebles (1984), Lahav et al. (1991), Eke, Cole \& Frenk (1996)
and Lokas \& Hoffman (2001). In the spherical collapse model, the
mass shell with a radius $R(t)$ contains a fixed mass $m$ and its
dynamics satisfies the energy equation
$$
\left(dR\over dt\right)^2 ={2Gm\over R} + {\Lambda R^2 \over 3} -
K  . \eqno(2)
$$
The constant $K$ is positive for a growing mass perturbation.
In a flat universe, the evolution of a scale factor is described by
$$
\left({1\over a}{da\over dt} \right)^2 = {8 \over 3} \pi G \rho_b
+ {\Lambda \over 3} , \eqno(3)
$$
where $\rho_b$ is the mean background density. Using eq. (3) we
can write eq. (2) as (Peebles 1984)
$$
\left(dR\over da\right)^2 = {a \over R} {(\omega R^3-\kappa R +1)
\over (\omega a^3 +1)} , \eqno(4)
$$
where $\kappa$ is a new constant, the present value of $a$ is
$a_0=1$ and
$$
\omega = {1 \over \Omega_0} -1. \eqno(5)
$$
The density contrast can be expressed as
$$
\delta={a^3\over R^3} - 1 . \eqno(6)
$$
The radial velocity in units of the Hubble velocity is
$$
{u\over HR} ={a\over R}{dR\over da}= 1 - {v\over HR}, \eqno(7)
$$
where $v$ is the peculiar velocity toward the centre of the system.

The perturbed region will turn around when
$$
\omega R^3 - \kappa R + 1 = 0. \eqno(8)
$$
Solving this cubic equation and requiring that a physically sensible root
exists, we find that the smallest perturbation which will
collapse has
$$
\kappa_{\rm min}={3 \omega^{1 \over 3} \over 2^{2\over 3}} .  \eqno(9)
$$
The corresponding turnaround radius is
$$
R_{\rm ta,max}= (2 \omega)^{-{1\over 3}} .\eqno(10)
$$
Perturbations with larger value of $\kappa$ turn around and
collapse earlier. Solving the cubic eq. (8) we can find the turnaround
radius $R_{ta}$ as a function of the density parameter $\omega$
and constant $\kappa$. The solution can be expressed as
(Eke, Cole \& Frenk 1996)
$$
R_{\rm ta}= - 2 s^{1\over 3}\cos\left({\theta \over 3} - {2\pi
\over 3}\right) ,\eqno(11)
$$
where
$$
s=\left(3\over 2\right)^3 \left(\kappa \over
\kappa^3_{\rm min}\right)^{3\over 2} \eqno(12)
$$
and $\theta$ satisfies
$$
\cos \theta =\left(\kappa_{\rm min} \over \kappa \right)^{{3 \over 2}}
 \,\,\,\,\,\,\,\,\,\,\, (0 < \theta < {\pi \over 2}) . \eqno(13)
$$

In the limit where $a \ll 1$, we find that
$$
\left({a \over R} {dR \over da} \right)^2 = {a^3 \over R^3}
(1-\kappa R) = 1+ \delta - \kappa a \eqno(14)
$$
and therefore, the density contrast can be written as
$$
\delta= - 2 \left(v \over HR \right) + \kappa a. \eqno(15)
$$
In linear perturbation theory the peculiar velocity for a spherical
mass concentration is determined as
$$
{v\over HR}={1\over 3} f(\Omega) \,\, \delta, \eqno(16)
$$
where $f(\Omega)$ is the dimensionless growth factor. For
$a \ll 1$, $f(\Omega)=1$. By using the relation (16), we find
from eq.(15) that
$$
\delta={3\kappa \over 5} a . \eqno(17)
$$
In the limit where $a \ll 1$, the density contrast is directly related
to the constant $\kappa$.

\begin{figure}
\centering \leavevmode \psfig{file=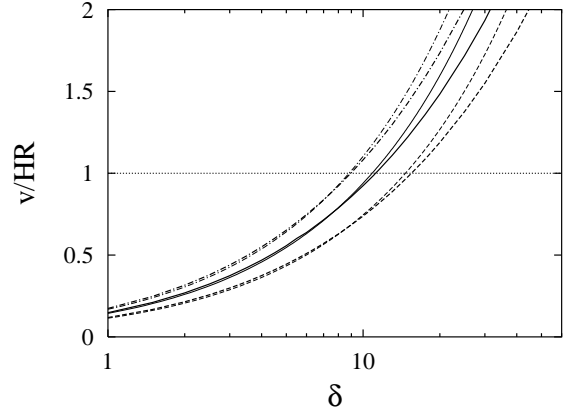,width=8cm}

\caption {Spherical collapse model. The peculiar velocity in units
of the Hubble velocity, $v/HR$, as a function of the density
contrast, $\delta$, in flat universes with the density parameter
$\Omega_0=0.2$ (dashed lines), $\Omega_0=0.3$ (solid lines) and
$\Omega_0=0.4$ (dot-dashed lines). Heavy lines show the exact
solutions obtained by integrating eq. (4) and light lines
represent the approximations given by eq. (18).}
\end{figure}

Fig.1 shows the relation between the peculiar velocity in units of
the Hubble velocity, $v/HR$, and the density contrast $\delta$ for
the flat models at the present moment. We integrated eq. (4) and
found the present density contrast and peculiar velocity from eq.
(6) and (7), respectively. For each value of the initial density
contrast $\delta_i$ at $a_i \ll 1$, the initial radius $R_i$ was
calculated using eq. (6) and the constant $\kappa$ was determined
using eq. (17). For comparison, we show the peculiar velocity
given by the approximation derived by Yahil (1985),
$$
{v\over HR}={1\over 3} f(\Omega_0) \, \delta \,
(1+\delta)^{-{1\over 4}}. \eqno(18)
$$
The growth factor $f(\Omega_0)$ is $0.41$, $0.51$ and $0.60$ for the
flat models with $\Omega_0=0.2,0.3$ and $0.4$, respectively.
Fig.~1 shows that the approximation (18) agrees well with the
numerical solution for the
density contrast $\delta<10$. For larger values of $\delta$, this
approximation overestimates the peculiar velocity.

Consider the density contrast at the turnaround point,
$\delta_{\rm ta}$, where $v=HR$. We found that $\delta_{\rm ta}=
15.48$, $11.19$ and $8.95$ in the flat models with $\Omega_0=0.2$,
$0.3$ and $0.4$, respectively. For comparison, we calculated also
the turnaround density contrast for the corresponding open models.
The turnaround density contrast in the open models can be
expressed as (Reg\"os \& Geller 1989)
$$
\delta_{\rm ta} = \left(\pi \over x_1 - \ln x_2\right)^2 \omega^3
-1, \eqno(19)
$$
where $x_1^2=(2\Omega_0^{-1}-1)^2-1$ and $x_2=x_1+2\Omega_0^{-1}-1$.
We found that $\delta_{\rm ta}=16.22$, $11.57$
and $9.17$ for the open models with $\Omega_0=0.2$, $0.3$ an $0.4$,
respectively. For the same value of $\Omega_0$, the turnaround density
in the open model is somewhat higher than in the flat model. However,
the effect of cosmological constant on $\delta_{\rm ta}$ is small.

Next we study the relation between the radial peculiar velocity and
density contrast for superclusters in N-body simulations.

\sec{NUMERICAL MODELS}
\ssec{Simulations}
First, we studied superclusters in a N-body simulation carried out
by the Virgo consortium for the flat $\Lambda$CDM model with a cosmological
constant. The Virgo simulations are described in detail by
Jenkins et al. (1998). These simulations were created using an
adaptive particle-particle/particle-mesh (AP$^3$M) code as
described by Couchman, Thomas \& Pearce (1995) and Pearce \&
Couchman (1997). In the $\Omega_0=0.3$ $\Lambda$CDM model studied
here, the power spectrum of the initial conditions was chosen
to be in the form given by Bond \& Efstathiou (1984),
$$
P(k)={Ak \over [1+(aq+(bq)^{3/2}+(cq)^2)^\nu]^{2/\nu}}, \eqno(20)
$$
where $q=k/\Gamma$, $a=6.4h^{-1}$Mpc, $b=3h^{-1}$Mpc,
$c=1.7h^{-1}$Mpc, $\nu=1.13$ and $\Gamma=\Omega_0 h = 0.21$. The
normalization constant, $A$, was chosen by fixing the value of
$\sigma_8$ (the linearly extrapolated mass fluctuation in spheres
of radius $8h^{-1}$Mpc) to be $0.9$. The evolution of particles
was followed in the comoving box of size $L=239.5h^{-1}$Mpc. The
number of particles was $N_p=256^3$. Therefore, the mean particle
separation $\lambda_p=L/N_p^{1/3}=0.9355 h^{-1}$Mpc. We denote
this $\Lambda$CDM model as the model M1.

We also investigated superclusters in a series of N-body
simulations we carried out for flat $\Lambda$CDM models, where the
density parameter $\Omega_0=0.2-0.4$ and $\sigma_8=0.7-0.9$. In
these simulations, the evolution of particles was followed by
using a particle-mesh (PM) code described by Gramann (1988) and
Suhhonenko \& Gramann (1999). The PM method is discussed in
detail by Hockney \& Eastwood (1981) and Efstathiou et al.
(1985). We used the PM code, where the number of cells was the
same as the number of particles. We denote these flat $\Lambda$CDM
models as the models M2-M3, S1-S4 and L1-L4.

In the models M2 and M3, the cosmological parameters were chosen
similar to the model M1. We chose the flat $\Omega_0=0.3$,
$\Omega_{\Lambda}=0.7$ model and used the initial power spectrum
$P(k)$ given in eq. (20) (with $\sigma_8=0.9$). In the model M2,
we followed the evolution of $256^3$ particles in the box of size
$L=239.5h^{-1}$Mpc. In the model M3, the number of particles was
$N_p=128^3$ and the box size $L=192h^{-1}$Mpc. Table~1 lists the
code, the number of particles and the box size used in models M1,
M2 and M3, respectively. Suhhonenko \& Gramann (2002) used the
models M1 and M2 to investigate the rms peculiar velocity of
galaxy clusters for different cluster masses and radii.

\begin{table}
\caption{N-body simulations with $\Omega_0=0.3$ and
$\sigma_8=0.9$. The power spectrum used is given in eq. (20). In
these simulations we studied superclusters in the density field
smoothed on the scale $R=5h^{-1}$Mpc.}

\begin{tabular}{|c|c|c|c|r|c|c|}
\hline \hline
 Model &  code  & $N_p$  &  $L$ ($h^{-1}$Mpc) & $N_s$ & $N_{0.5}$ & $N_c$  \\

\hline

 M1    & AP$^3$M & 256$^3$ & 239.5 & 1414 & 342 & 63 \\
 M2    & PM      & 256$^3$ & 239.5 & 1374 & 259 & 23 \\
 M3    & PM      & 128$^3$ & 192   &  662 & 153 & 14 \\

\hline
\end{tabular}
\label{table}
\end{table}

In the models S1-S4 and L1-L4, the number of particles was
$N_p=128^3$. In these models, the initial power spectrum was
calculated by using the fast Boltzmann code CMBFAST developed by
Seljak \& Zaldarriaga (1996). The normalized Hubble constant was
chosen to be $h=0.7$ and the baryon density $\Omega_{b}
h^2=0.02$. Table~2 lists the density parameter $\Omega_0$,
$\sigma_8$ and the box size $L$ used in the models S1-S4 and
L1-L4. We studied the flat models with a cosmological constant
$\Omega_{\Lambda}=1-\Omega_0$.

\begin{table}
\caption{N-body simulations, where the evolution of $128^3$
particles was followed by the PM code. The power spectrum of the
initial density field was calculated by using the code CMBFAST.
In the models S1-S4, we studied superclusters in the density field
smoothed on the scale $R=5h^{-1}$Mpc. In the models L1-L4, we used
the smoothing scale $R=10h^{-1}$Mpc.}

\begin{tabular}{|c|c|c|c|c|r|r|}
\hline \hline
 Model & $\Omega_0$ & $\sigma_8$ & $L$ ($h^{-1}$Mpc)& $N_s$ & $N_{0.5}$
& $N_c$  \\

\hline

 S1    & 0.2 & 0.8 & 192 & 541 &  80 &  4  \\
 S2    & 0.3 & 0.7 & 192 & 633 & 119 & 11  \\
 S3    & 0.3 & 0.9 & 192 & 630 & 149 & 20 \\
 S4    & 0.4 & 0.8 & 192 & 699 & 191 & 25 \\

\hline
 L1    & 0.2 & 0.8 & 384 & 606 & 20 & --- \\
 L2    & 0.3 & 0.7 & 384 & 611 & 14 & --- \\
 L3    & 0.3 & 0.9 & 384 & 687 & 57 & --- \\
 L4    & 0.4 & 0.8 & 384 & 718 & 52 & --- \\
\hline
\end{tabular}
\label{table}
\end{table}

For the models M3 and S3, all the parameters are the same, except of
the form of the initial power spectrum. In the model M3, we used
the approximation given in eq. (20) and in the model S3, the power
spectrum was calculated numerically by using the code CMBFAST.

\begin{figure}
\centering
\leavevmode
\psfig{file=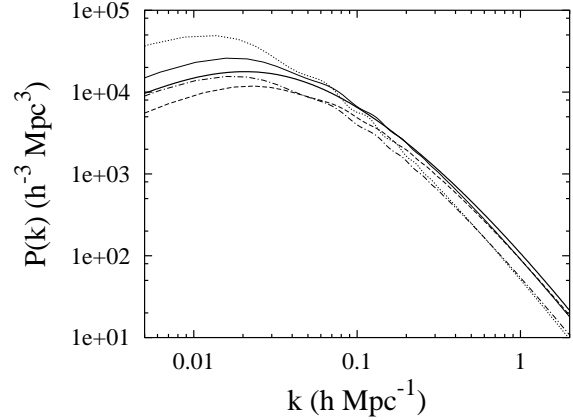,width=8cm}

\caption {Power spectra used in the N-body simulations. The heavy
solid line shows the power spectrum given in eq. (20) for
$\Omega_0=0.3$ and $\sigma_8=0.9$. The light curves show the
power spectra calculated by using the code CMBFAST. The dotted
line shows the power spectrum for $\Omega_0=0.2$, $\sigma_8=0.8$,
the solid line for $\Omega_0=0.3$, $\sigma_8=0.9$, the dot-dashed
line for $\Omega_0=0.3$, $\sigma_8=0.7$ and the dashed line for
$\Omega_0=0.4$, $\sigma_8=0.8$. The normalized Hubble constant
$h=0.7$.}

\end{figure}

Fig.~2 shows the power spectra used in the N-body simulations for
different values of the parameters $\Omega_0$ and $\sigma_8$.
The light curves show the numerical power spectra calculated by
using the code CMBFAST. In the models studied, the power spectrum
on large scales is quite different. For the same value of
$\sigma_8$, the amplitude of the large-scale density fluctuations
is highest for $\Omega_0=0.2$ and lowest for $\Omega_0=0.4$. The
light solid line shows the power spectrum for $\Omega_0=0.3$ and
$\sigma_8=0.9$. For comparison, we show the power spectrum given
in eq. (20) for $\Omega_0=0.3$ and $\sigma_8=0.9$ (heavy solid
line). We see that on large scales, the numerical power spectrum
is considerably higher than the power spectrum given in eq. (20)
($\sim 26$ \% at the wavelength $\lambda=2\pi/k=192h^{-1}$Mpc).

\ssec{Selection of superclusters}
To select the superclusters in the N-body simulations, we used the
following method:

(1) We calculated the density contrast on a grid. For each grid
point, the density contrast was determined as
$$
\delta = {N \over \bar N} -1,
\eqno(21)
$$
where $N$ is the number of particles in the sphere of radius $R$ around
the grid point, and
$$
\bar N = {4 \pi \over 3} {N_p \, R^3 \over L^3}
\eqno(22)
$$
is the mean number of particles in the sphere of radius $R$. We
denote the number of cells in the grid as $N_g^3$.

(2) We found the density maxima on the grid. The grid point was
considered as a density maximum, if its density contrast was
higher than the density contrast in all neighbouring grid points.
In the following, we studied the density maxima, where the
density contrast was $\delta>1$.

(3) We identified the candidate supercluster centres. For each
grid point, where the density maximum was determined, we found all
particles in a cube with the size $l=L/N_g$ around
this grid point. For each particle we calculated the
density contrast using eq. (21), where $N$ was the number
of particles in the sphere of radius $R$ around this fixed
particle. The location of the particle, where the density
contrast had the maximum value, was identified as the candidate
supercluster centre.

(4) The final supercluster list was obtained by deleting the
candidate superclusters with lower density contrast in all pairs
separated by less than the radius $R$.

In this way we define the superclusters as maxima of the density
field smoothed with a top-hat window with radius $R$. In the
models M1-M3 and S1-S4, we used the smoothing radius
$R=5h^{-1}$Mpc. In the models M1 and M2, the mean number of
particles in spheres of $R=5h^{-1}$Mpc is $\bar N=639.4$, in the
models M3 and S1-S4 - $\bar N=155.1$. To select superclusters for
the models M1 and M2, we used a $80^3$ grid. In this case, the
cell size was $l=L/N_g=2.99h^{-1}$Mpc. To identify superclusters
for the models M3 and S1-S4, we used a $64^3$ grid
($l=3h^{-1}$Mpc). In the models L1-L4, we studied the
superclusters defined with $R=10h^{-1}$Mpc. In these models, the
mean number of particles in spheres of $R=10h^{-1}$Mpc is $\bar
N=155.1$. To select superclusters, we used a $64^3$ grid
($l=6h^{-1}$Mpc).

\sec{RADIAL VELOCITY AND DENSITY CONTRAST IN THE SUPERCLUSTERS}
Table~1 and Table~2 show the number of superclusters, $N_s$,
identified in different models. For each supercluster, we
investigated the density contrast, $\delta$, and the radially
averaged peculiar velocity, $v$. The density contrast in a
supercluster was determined as
$$
\delta = {N_f \over \bar N} -1,
\eqno(23)
$$
where $N_f$ is the number of particles in the sphere of
radius $R$ around the centre of the supercluster. The radially averaged
peculiar velocity for the supercluster, in a shell of radius $R$,
was defined as
$$
v= - {1\over N_l} \, {\sum_{i=1}^{N_l} (\vec v_{i}^{\,\, \prime} -
\vec v_{0}^{\,\, \prime}) \,\cdot \, \hat{\vec r_i} } , \eqno(24)
$$
where $N_l$ is the number of particles in the shell $0.1R$ wide,
$\vec v_i^{\,\, \prime}$ is the peculiar velocity of the particle
$i$ at a radial distance $\vec r_i$ from the centre of the
supercluster, $\hat{\vec r_i}$ is the unit vector in the
direction of the particle $i$ and $\vec v_{0}^{\,\, \prime}$ is the
mean peculiar velocity of all the particles within the sphere of
radius $R$,
$$
\vec v_{0}^{\,\, \prime}= {1\over N_f} \, {\sum_{i=1}^{N_f} \vec
v_i^{\,\, \prime}}. \eqno(25)
$$

\begin{figure*}
\centering \leavevmode \psfig{file=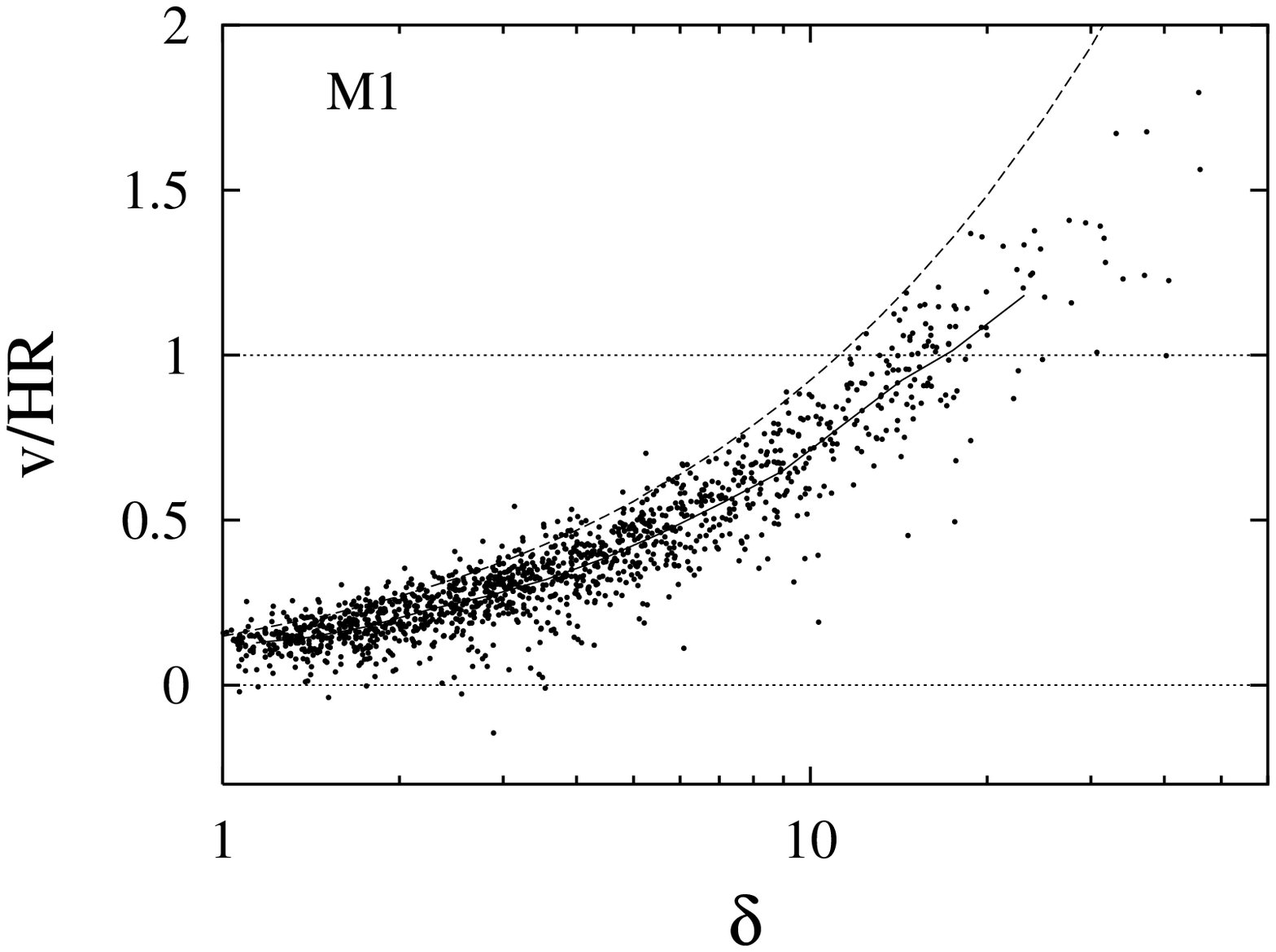,width=8cm}
\psfig{file=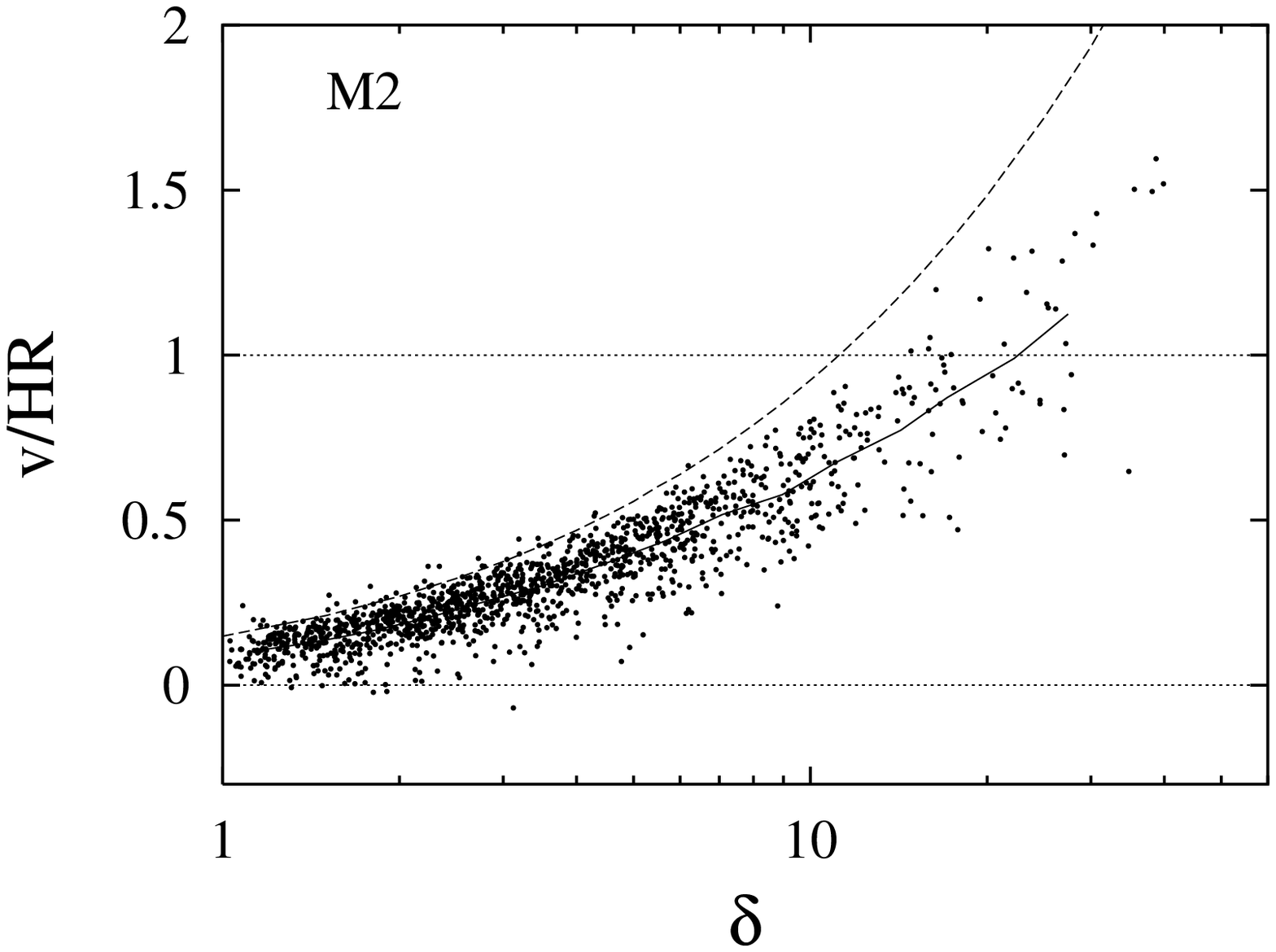,width=8cm} \psfig{file=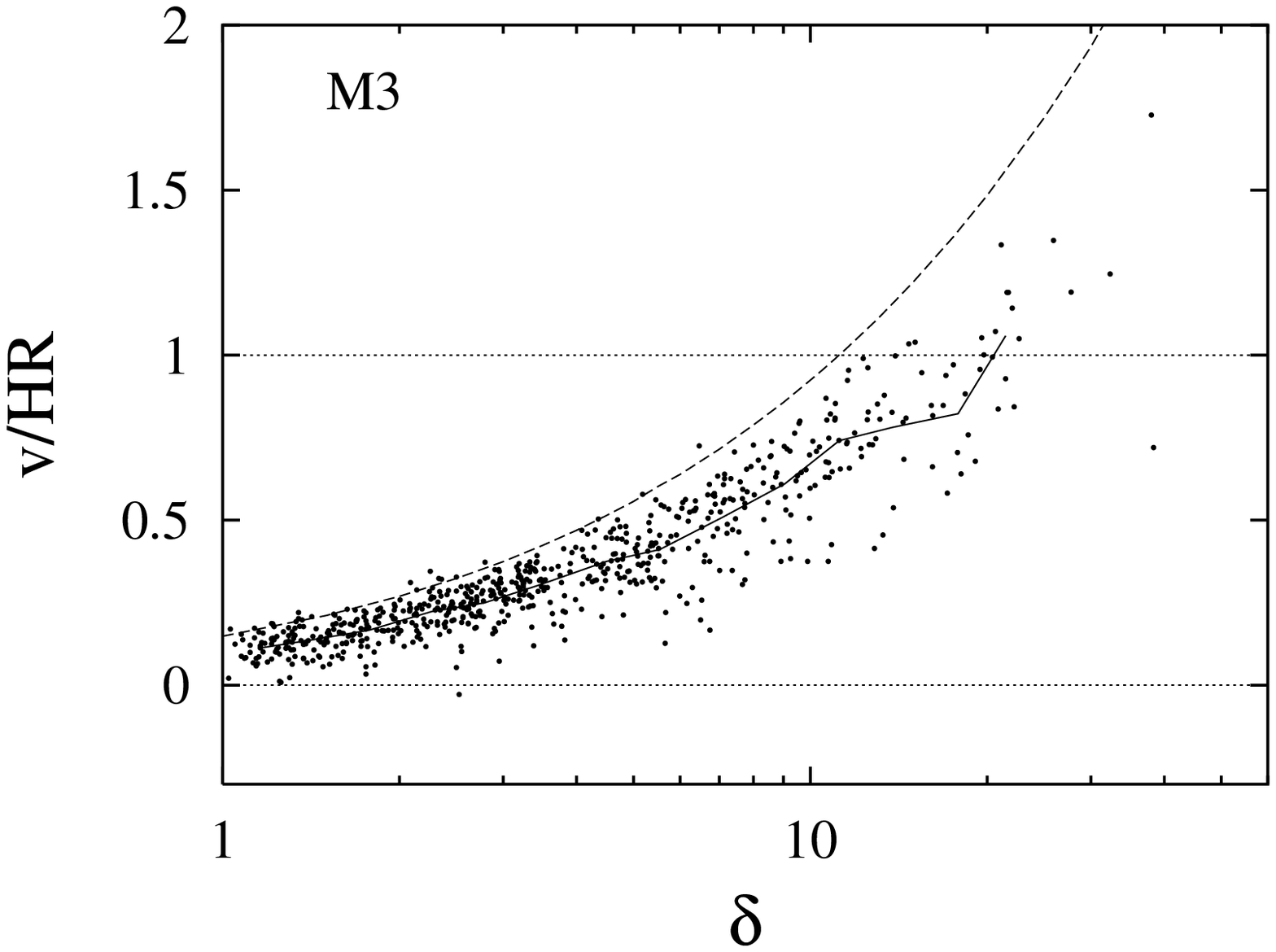,width=8cm}
\psfig{file=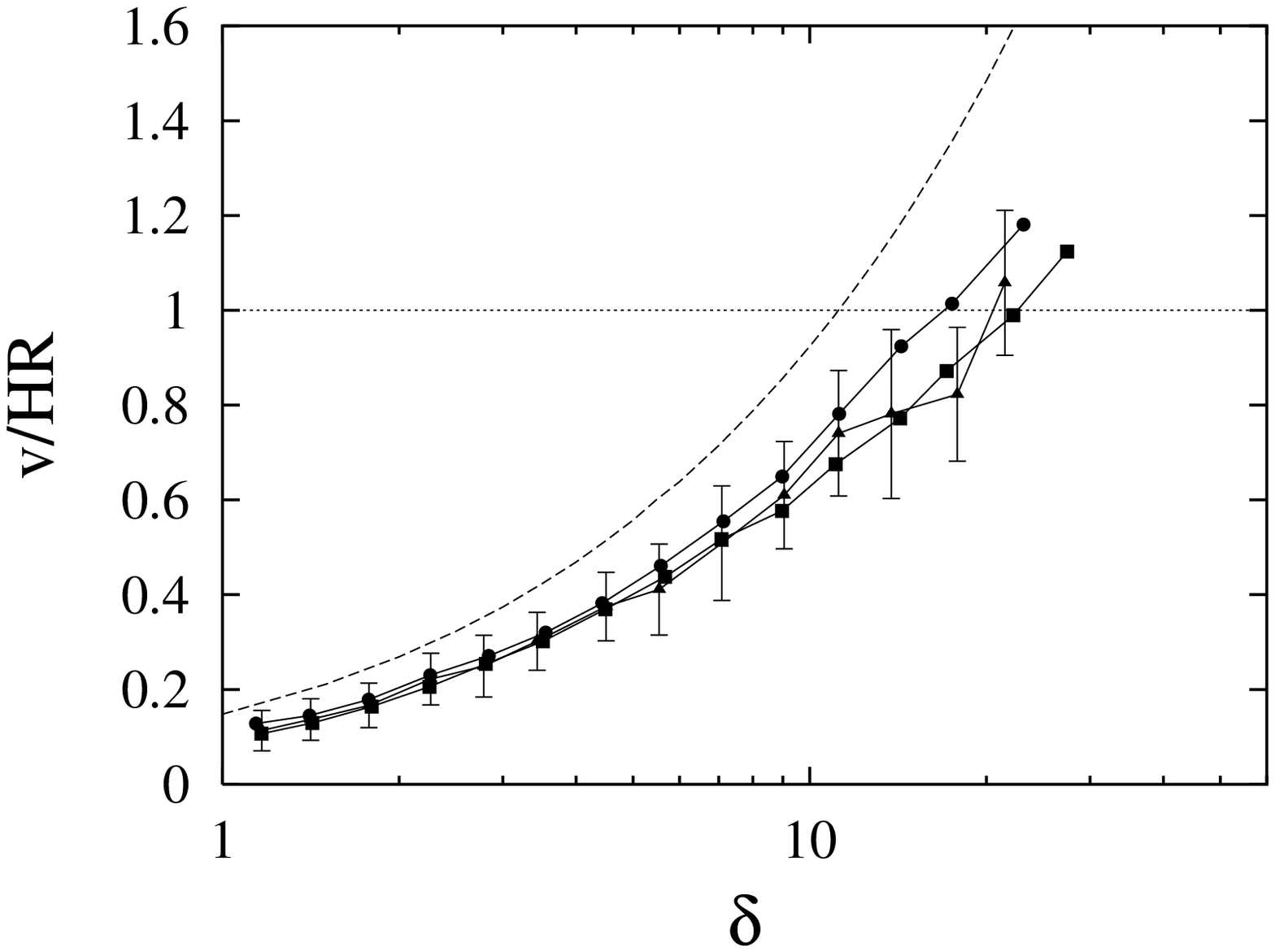,width=8cm}

\caption{The averaged radial peculiar velocity of a supercluster
in units of the Hubble velocity, $v/HR$, vs. the density contrast
of the supercluster in the models M1 (upper right panel), M2
(upper left panel) and M3 (lower left panel). The solid lines
show the mean peculiar velocity for the superclusters. For
comparison, we show the peculiar velocity in the spherical
collapse model for $\Omega_0=0.3$ (dashed lines). The lower right
panel shows the mean peculiar velocity in the models M1
(circles), M2 (squares) and M3 (triangles). The error bars show
the standard deviation. For clarity, error bars are only shown
for the model M3. The superclusters are defined with
$R=5h^{-1}$Mpc.}

\end{figure*}

Fig.~3 shows the averaged radial peculiar velocity of a
supercluster in units of the Hubble velocity, $v/HR$, vs. the
density contrast of the supercluster in the models M1, M2 and M3.
The superclusters were defined with $R=5h^{-1}$Mpc. In the models
M1, M2 and M3, the number of superclusters is $1414$, $1374$ and
$662$, respectively. We see a close correlation between the
peculiar velocity and the density contrast of a supercluster. We
can estimate the dynamical state of the supercluster on the basis
of its density contrast.

For comparison, we also show in Fig.~3 the peculiar velocity
predicted in the flat spherical collapse model for
$\Omega_0=0.3$. The radially averaged peculiar velocity of
superclusters is systematically smaller than that predicted by the
spherical collapse model.

We investigated the mean peculiar velocity and the standard
deviation for the superclusters in different density
intervals. The superclusters were divided into subgroups
according to their density contrast. We studied the velocities in
fifteen subgroups, where the logarithm of the density contrast
was in the range $\log \delta = 0 - 0.1$, $0.1 - 0.2$, ..., $1.4 - 1.5$.
The mean peculiar velocity and standard deviation were determined for
the intervals which contained at least eight superclusters. The
lower right panel in Fig.~3 shows the mean peculiar velocity
determined in the models M1, M2 and M3. The error bars show the
standard deviation. For clarity, the standard deviation is only
shown for the model M3. In the models M1, M2 and M3, the standard
deviation is similar.

The models M1 and M2 differ from each other by the code used to
follow the evolution of particles. The model M1 was created using
a AP$^3$M code and the model M2 by using a PM code. Fig. 3 shows
that in the model M1, the mean peculiar velocities are somewhat
higher than in the model M2. For example, in the range
$\delta=10.0 - 12.6$, the mean peculiar velocities are $\bar v=0.78 HR$ and
$\bar v=0.68 HR$ in the models M1 and M2, respectively. This difference
is probably due to the smoothing inherent to the PM method.
However, the effect of the code on the mean peculiar velocity is
not large ($\sim 15$\% in the range $\delta=10.0 - 12.6$).

In the model M3, the box size $L$ and the number of particles,
$N_p$, are smaller than in the models M1 and M2. In this model for
$R=5h^{-1}$Mpc, the box size $L=38.4R$ and the mean particle
separation $\lambda_p=0.3 R$. The same values for $L/R$ and
$\lambda_p/R$ are used in the models S1-S4 and L1-L4. Fig.3 shows
that the mean peculiar velocities in the model M3 are similar to
the mean velocities in the models M1 and M2. For example, in the
range $\delta=10-12.6$, the mean peculiar velocity is $\bar
v=0.74 HR$. Therefore, a box size $L=38.4 R$ and a mean particle
separation $\lambda_p=0.3 R$ are sufficient to study the mean
peculiar velocities in superclusters defined with radius $R$.

We can compare the mean peculiar velocities of superclusters
with the predictions of the spherical collapse model. Lets
consider the model M1. In this model, for the range
$\delta=1.00-1.26$, the mean density contrast is $1.14$ and the
mean peculiar velocity is $\bar v=0.13 HR$. For comparison, in
the spherical collapse model, the peculiar velocity for
$\delta=1.14$ is $v=0.17 HR$. For the range $\delta=10.0-12.6$,
the mean density contrast for superclusters is $11.2$ and
$\bar v=0.78 HR$. In the spherical collapse model, the peculiar
velocity for $\delta=11.2$ is $v=1.0 HR$. Therefore, the mean
peculiar velocities in the simulation are $\sim 25\%$ smaller
than those predicted by the spherical collapse model. The
difference in the peculiar velocities for $\delta \approx 1$ and
$\delta \approx 10$ is similar. In the model M2 and M3, the deviations
from the spherical collapse model are somewhat larger than in the
model M1 ($\sim 35$\% and $\sim 30$\% for the model M2 and M3,
respectively).

\begin{figure}
\centering
\leavevmode
\psfig{file=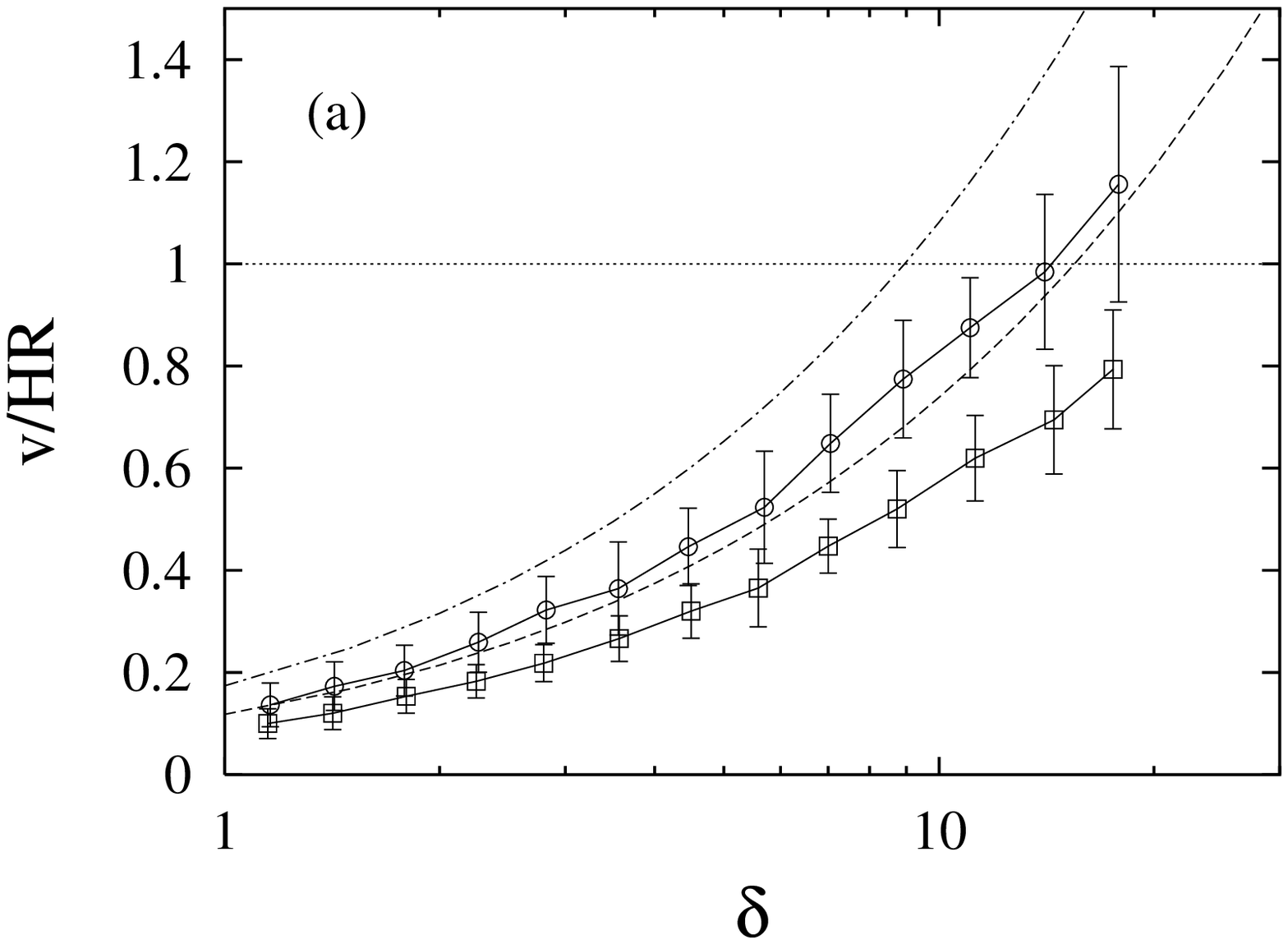,width=8cm}
\psfig{file=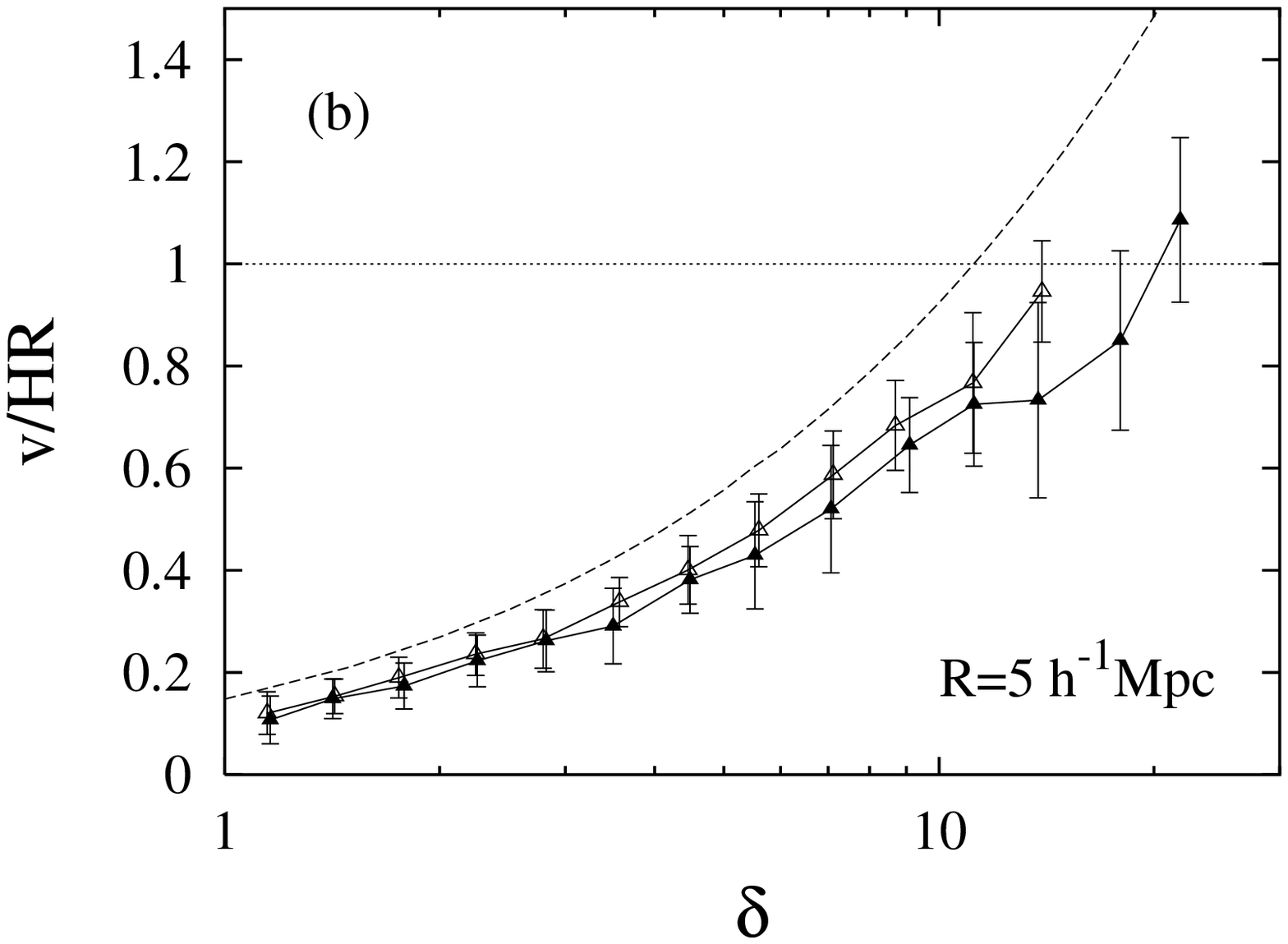,width=8cm}

\caption{The radially averaged peculiar velocity for the superclusters in
the models S1-S4. The superclusters are defined
in the density field smoothed on the scale $R=5h^{-1}$Mpc.
Panel (a) shows the peculiar velocities for the model S1 (squares)
and S4 (circles). For comparison, we show the peculiar velocities
predicted by the spherical collapse model for $\Omega_0=0.2$
(dashed line) and for $\Omega_0=0.4$ (dot-dashed line). Panel (b)
shows the peculiar velocities for the model S2 (open triangles) and
S3 (filled triangles). The dashed line describes the peculiar
velocities in the spherical collapse model for $\Omega_0=0.3$.}

\end{figure}

Fig.~4 shows the radially averaged peculiar velocities for the
superclusters in the models S1-S4. The mean peculiar velocity with
a standard deviation is determined similarly as in the models M1-M3.
Panel (a) shows the peculiar velocities for the model S1 ($\Omega_0=0.2$)
and S4 ($\Omega_0=0.4$). For the same value of the density contrast, the
mean peculiar velocity is larger, if the density parameter
$\Omega_0$ is larger. For the range $\delta=5.0-6.3$, the
peculiar velocity $v/HR=0.37 \pm 0.08$ and $v/HR=0.52 \pm 0.11$
in the model S1 and S4, respectively. For the range
$\delta=10-12.6$, the peculiar velocity $v/HR=0.62 \pm 0.08$ and
$v/HR=0.87 \pm 0.10$, respectively.

Panel (b) in Fig.~4 shows the peculiar velocities for the superclusters
in the model S2 and S3. In these models $\Omega_0=0.3$. In the model
S2, the parameter $\sigma_8=0.7$ and in the model S3, the
$\sigma_8=0.9$. We see that the mean peculiar velocities in the model
S3 are somewhat smaller than the peculiar velocities in the model S2.
The difference between the models S2 and S3 is not large.
For the range $\delta=5.0-6.3$, the peculiar velocity
$v/HR=0.48 \pm 0.07$ and $v/HR=0.43 \pm 0.10$ in the model S2 and S3,
respectively. For the range $\delta=10-12.6$, the peculiar velocity
$v/HR=0.77 \pm 0.14$ and $v/HR=0.73 \pm 0.12$, respectively.
The peculiar velocities in the model S3 are similar to the peculiar
velocities in the model M3.

We compared the mean peculiar velocities for the superclusters in
the models S1-S4 with the predictions of the spherical collapse
model. The mean peculiar velocities in the models S1, S2 and S4
are $\sim 25$\% smaller than those predicted by the spherical collapse
model. In the model S3, the difference is  $\sim 30$\%. In all
models, this difference is similar for different values of the
density contrast.

We also determined the turnaround density contrast $\delta_{\rm
ta}$ for the $R=5h^{-1}$Mpc superclusters. We found all the
superclusters, where the peculiar velocity in units of the Hubble
velocity, $v/HR$, is in the range [0.9, 1.1] and calculated the
mean density contrast and the standard deviation for these
superclusters. Table~3 shows the density contrast $\delta_{\rm ta}$ for
the models studied. For the model M1, we found that $\delta_{\rm
ta}=16.2\pm 4.6$. For comparison, in the spherical collapse model the
turnaround density contrast is $11.19$ for $\Omega_0=0.3$ (see Section 2).
For the model S1 and S4, we found that $\delta_{\rm ta}=23.7 \pm 7.1$
and $\delta_{\rm ta}=12.3 \pm 2.2$, respectively. In the
spherical model, $\delta_{ta}=15.48$ and $8.95$ for
$\Omega_0=0.2$ and $\Omega_0=0.4$, respectively. The turnaround
density contrast for the superclusters in N-body simulations is
substantially larger than the turnaround density contrast in the
spherical collapse model.

\begin{table}
\caption{The turnaround density contrast, $\delta_{\rm ta}$, for
the superclusters studied with $R=5h^{-1}$Mpc.}

\begin{tabular}{|c|c|}
\hline \hline
 Model & $\delta_{\rm ta}$  \\

\hline

 M1    & $16.2 \pm 4.6$ \\
 M2    & $18.1 \pm 4.3$ \\
 M3    & $16.8 \pm 3.6$  \\

\hline

 S1    & $23.7 \pm 7.1$   \\
 S2    & $14.6 \pm 1.8$   \\
 S3    & $18.2 \pm 4.0$    \\
 S4    & $12.3 \pm 2.2$   \\

\hline

\end{tabular}
\label{table}
\end{table}

\begin{figure}
\centering
\leavevmode
\psfig{file=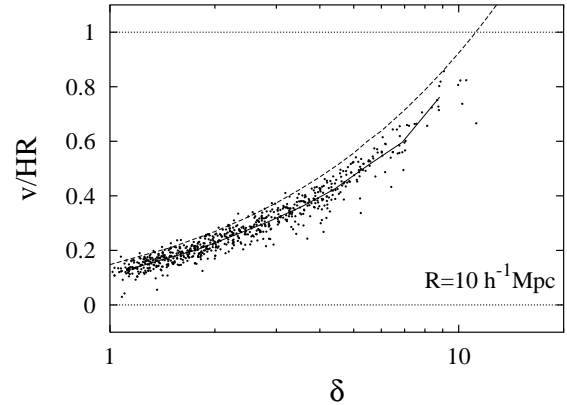,width=8cm}

\caption{The radially averaged peculiar velocity for 687
superclusters of the model L3. The superclusters are defined with
$R=10h^{-1}$ Mpc. The solid line shows the mean peculiar velocity
for the superclusters and the dashed line describes the peculiar
velocity in the spherical collapse model for $\Omega_0=0.3$.}

\end{figure}

We also studied superclusters in the models L1-L4. In these
models the box size was $L=384h^{-1}$Mpc and we investigated the
superclusters in the density field smoothed on the scale
$R=10h^{-1}$Mpc. Fig.~5 shows the radially averaged peculiar
velocity of a supercluster in units of the Hubble velocity,
$v/HR$, vs. the density contrast of the supercluster, for 687
superclusters of the model L3. For comparison, we show the
peculiar velocities predicted in the spherical collapse model for
$\Omega_0=0.3$. We see that also for the superclusters studied
with $10h^{-1}$Mpc smoothing, the mean infall velocities are
systematically smaller than those predicted by the spherical
collapse model.

Fig.~6 shows the peculiar velocities for the superclusters in the models
L1-L4. For comparison, the dotted line in panel (b) shows the peculiar
velocities for the $R=5h^{-1}$Mpc superclusters in the model M1. For the
$R=10h^{-1}$Mpc superclusters, the mean peculiar velocities in
units of the Hubble velocity, $\bar v/HR$, are somewhat higher
than for the $R=5h^{-1}$Mpc superclusters. For the range
$\delta=5.0-6.3$, the peculiar velocity $v/HR=0.43 \pm 0.04$ and
$v/HR=0.63 \pm 0.06$ in the model L1 and L4, respectively. The
peculiar velocity $v/HR= 0.54 \pm 0.04$ and $v/HR=0.51 \pm 0.05$, in
the model L2 and L3, respectively. The mean peculiar velocities for the
superclusters in the models L1-L4 are $\sim 15$\% smaller than
those predicted by the spherical collapse model.

\begin{figure}
\centering
\leavevmode
\psfig{file=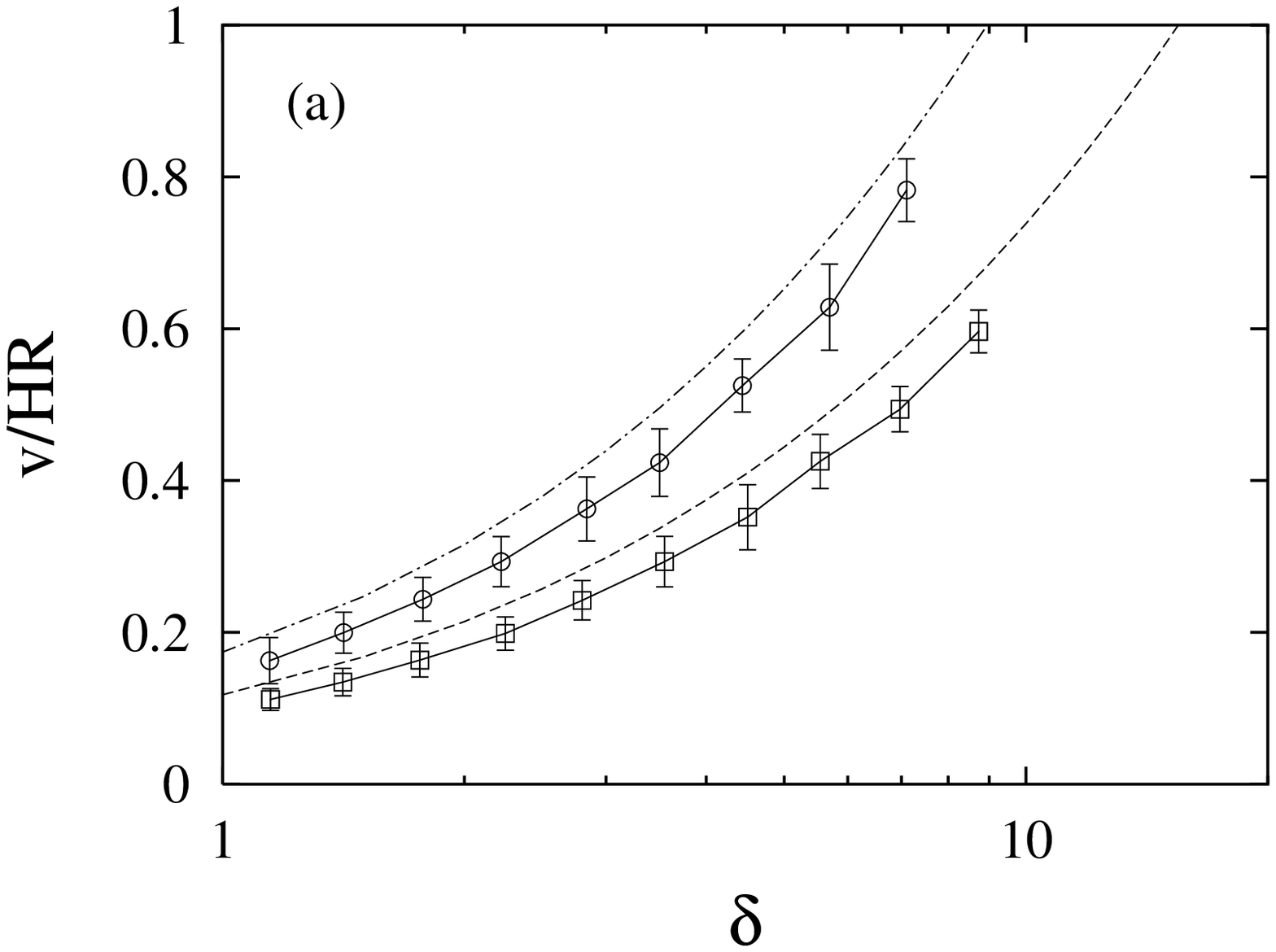,width=8cm}
\psfig{file=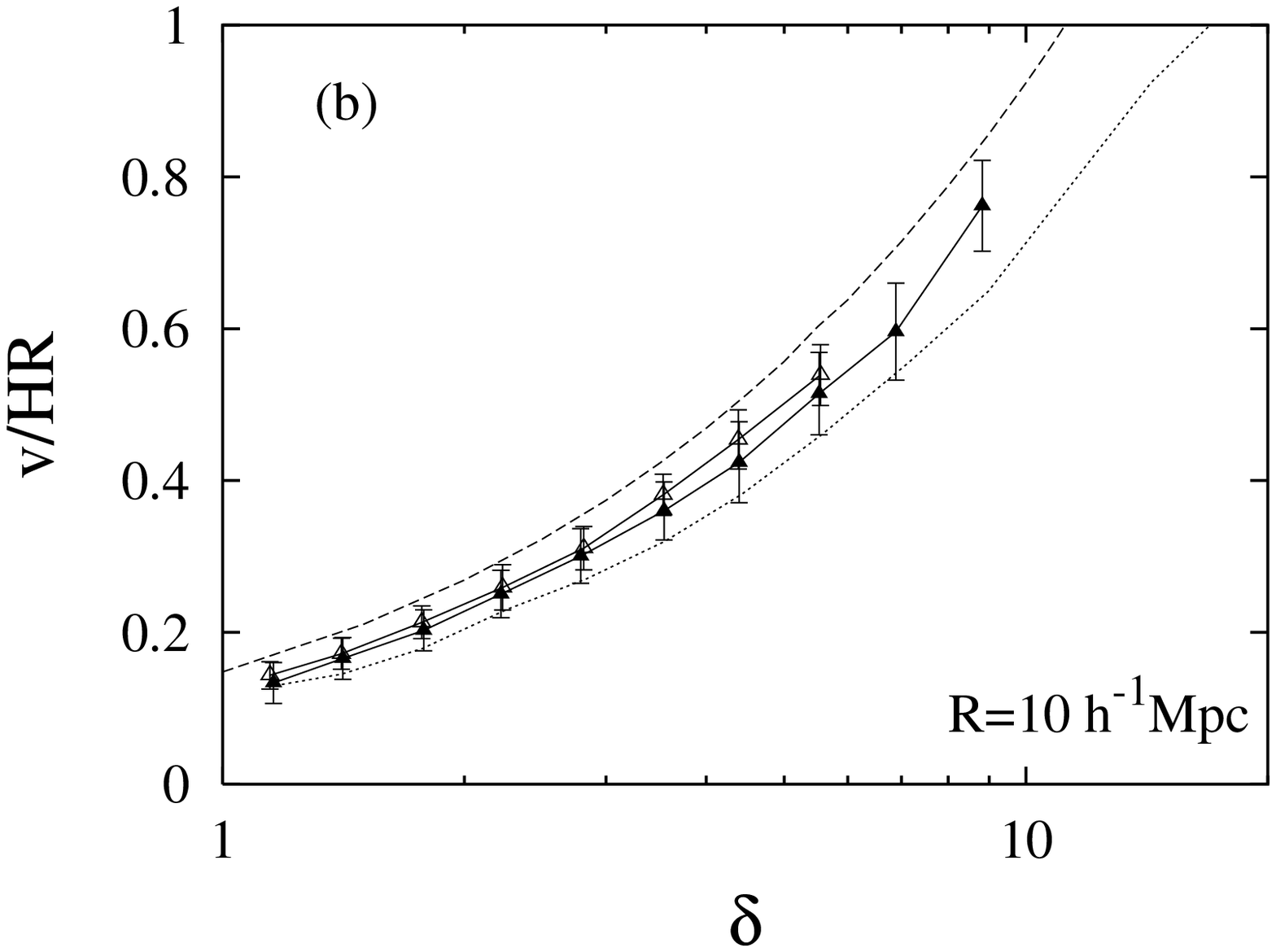,width=8cm}

\caption{The peculiar velocity for the superclusters in the
models L1-L4.  Panel (a) shows the peculiar velocities for the
model L1 (squares) and L4 (circles). For comparison, we show the
peculiar velocities predicted by the spherical collapse model for
$\Omega_0=0.2$ (dashed line) and for $\Omega_0=0.4$ (dot-dashed
line). Panel (b) shows the peculiar velocities for the model L2
(open triangles) and L3 (filled triangles). The dotted line shows
the mean peculiar velocity for the $R=5h^{-1}$Mpc superclusters
in the model M1 and the dashed line describes the peculiar
velocities in the spherical collapse model for $\Omega_0=0.3$.}

\end{figure}

\sec{VELOCITY DISTRIBUTION FOR THE SUPERCLUSTERS}

We investigated the number of superclusters, where the radially
averaged peculiar velocity is larger than a given value, $N(>v)$.
We denote the number of superclusters, which have peculiar
velocities $v>HR$ and, therefore, have already started to collapse,
as $N_c$. We also examined the number of superclusters, which have
peculiar velocities $v>0.5 HR$. We denote this number as $N_{0.5}$.
Table~1 and Table~2 show the numbers $N_c$ and $N_{0.5}$ for the
different models studied. In the model M1, we found $63$ superclusters
which have already started to collapse.

Panel (a) in Fig.~7 shows the number density of superclusters, where the
radially averaged peculiar velocity is larger than a given value, $n(>v)$.
The number density is given for the superclusters in the models M1-M3. The
function $n(>v)$ was determined as
$$
n(>v)={N(>v) \over L^3}.
\eqno(26)
$$
We also show the Poisson error bars for the number
densities. Panel (b) in Fig.~7 demonstrates the density
distribution for the superclusters in the models M1-M3. The
density distribution in these models is similar, but the velocity
distribution is different. The function $n(>v)$ for the model M1
is higher than for the model M2. In the model M1 and M2, the
number density of superclusters that have peculiar velocities
$v>HR$, is $4.6 \times 10^{-6}h^{3}$Mpc$^{-3}$ and $1.7 \times
10^{-6} h^3$Mpc$^{-3}$, respectively. The difference between the
models M1 and M2 is smaller for smaller peculiar velocities. The
velocity distribution in the model M3 is similar to the model M2.
Therefore, the number density $n(>v)$ in the model M3 is probably
underestimated.

\begin{figure}
\centering
\leavevmode
\psfig{file=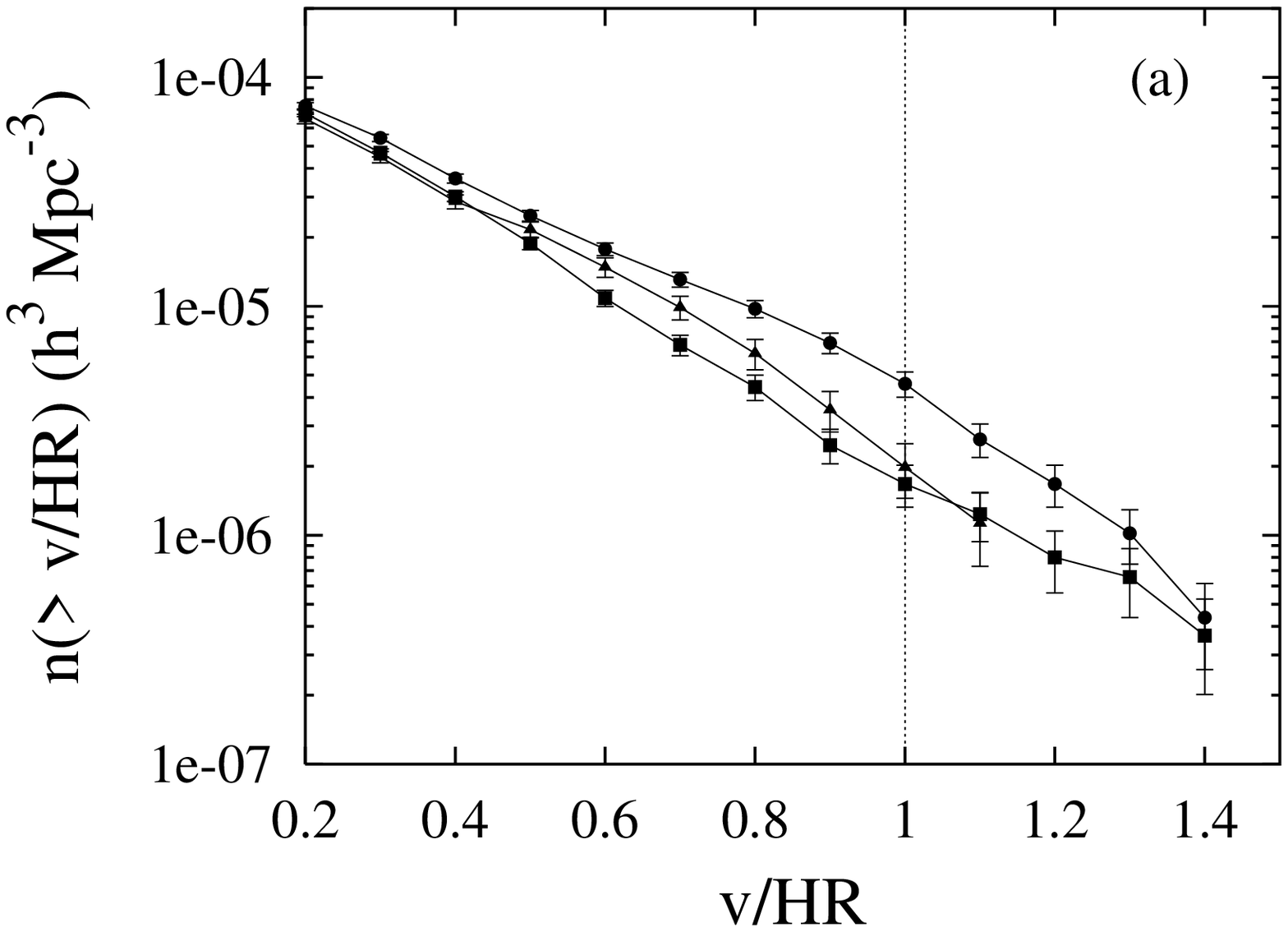,width=8cm}
\psfig{file=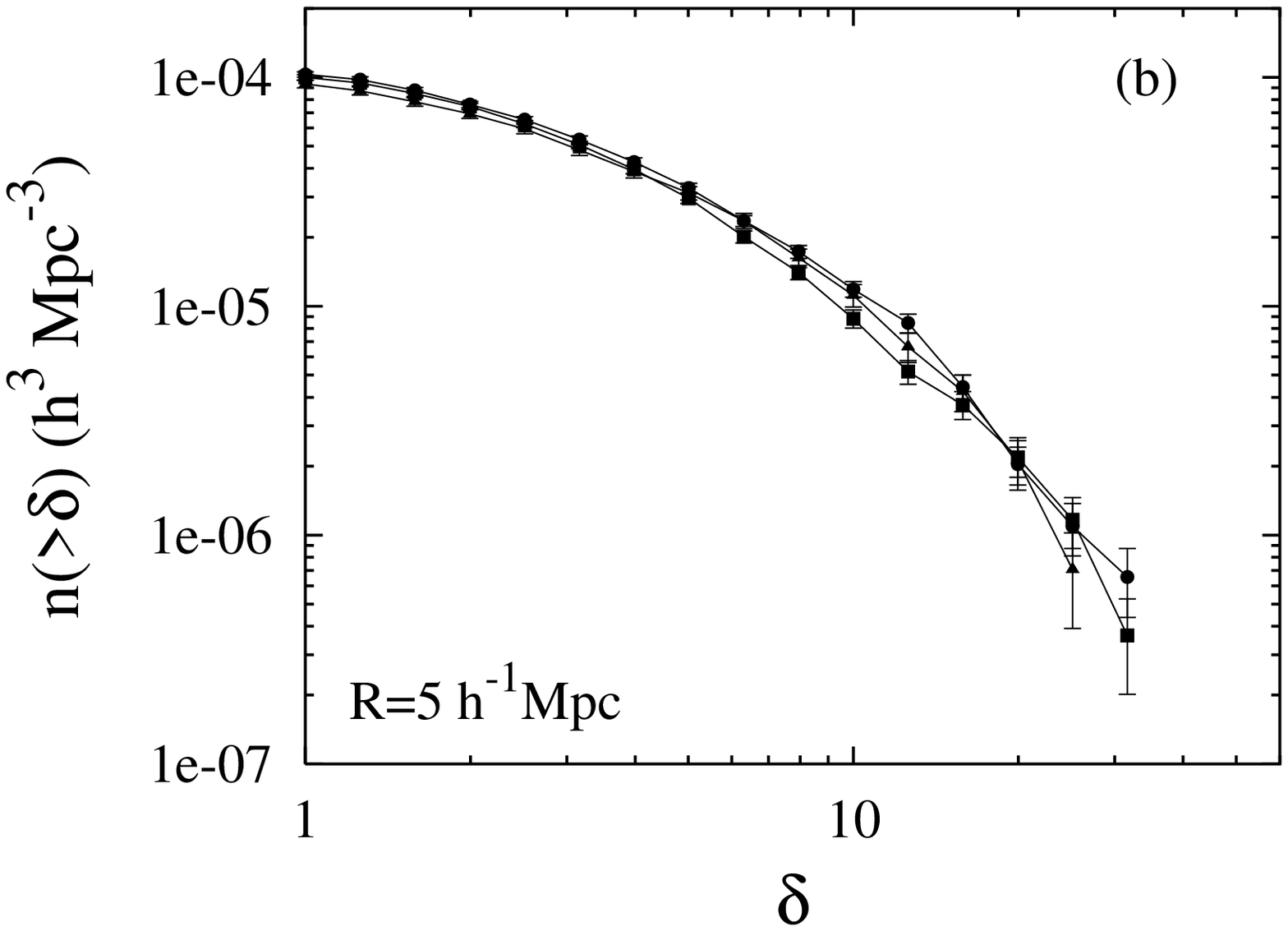,width=8cm}

\caption{(a) The number density of superclusters, where the radially
averaged peculiar velocity is larger than a given value, $n(>v)$.
The number density is shown for the superclusters in the model M1 (circles),
M2 (squares) and M3 (triangles). (b) The number density of superclusters,
where the density contrast is larger than a given value, $n(>\delta)$.
The density distribution is given for the same models as in panel a.
Error bars in panels a and b denote Poisson errors.}

\end{figure}

Fig.~8 shows the number density, $n(>v)$, for the superclusters in the
models S1-S4 and L1-L4. We can use models S1-S4 and L1-L4 to estimate the
lower limit for the function $n(>v)$ in different models. Also, we
can examine the relative differences between the different models.
Panel (a) in Fig.~8 shows the number density for the $R=5h^{-1}$Mpc
superclusters in the models S1-S4 ($L=192h^{-1}$Mpc). The function
$n(>v)$ is lowest for the model S1 ($\Omega_0=0.2$) and highest for the
model S4 ($\Omega_0=0.4$). In the model S1, we found $4$ superclusters,
which have started to collapse. In the model S4, this number was $25$. The
number density of superclusters, which have peculiar velocities $v>0.5 HR$,
is $1.1 \times 10^{-5}h^3$Mpc$^{-3}$ and $2.7 \times 10^{-5}h^{3}$Mpc$^{-3}$,
in the model S1 and S4, respectively.

\begin{figure}
\centering
\leavevmode
\psfig{file=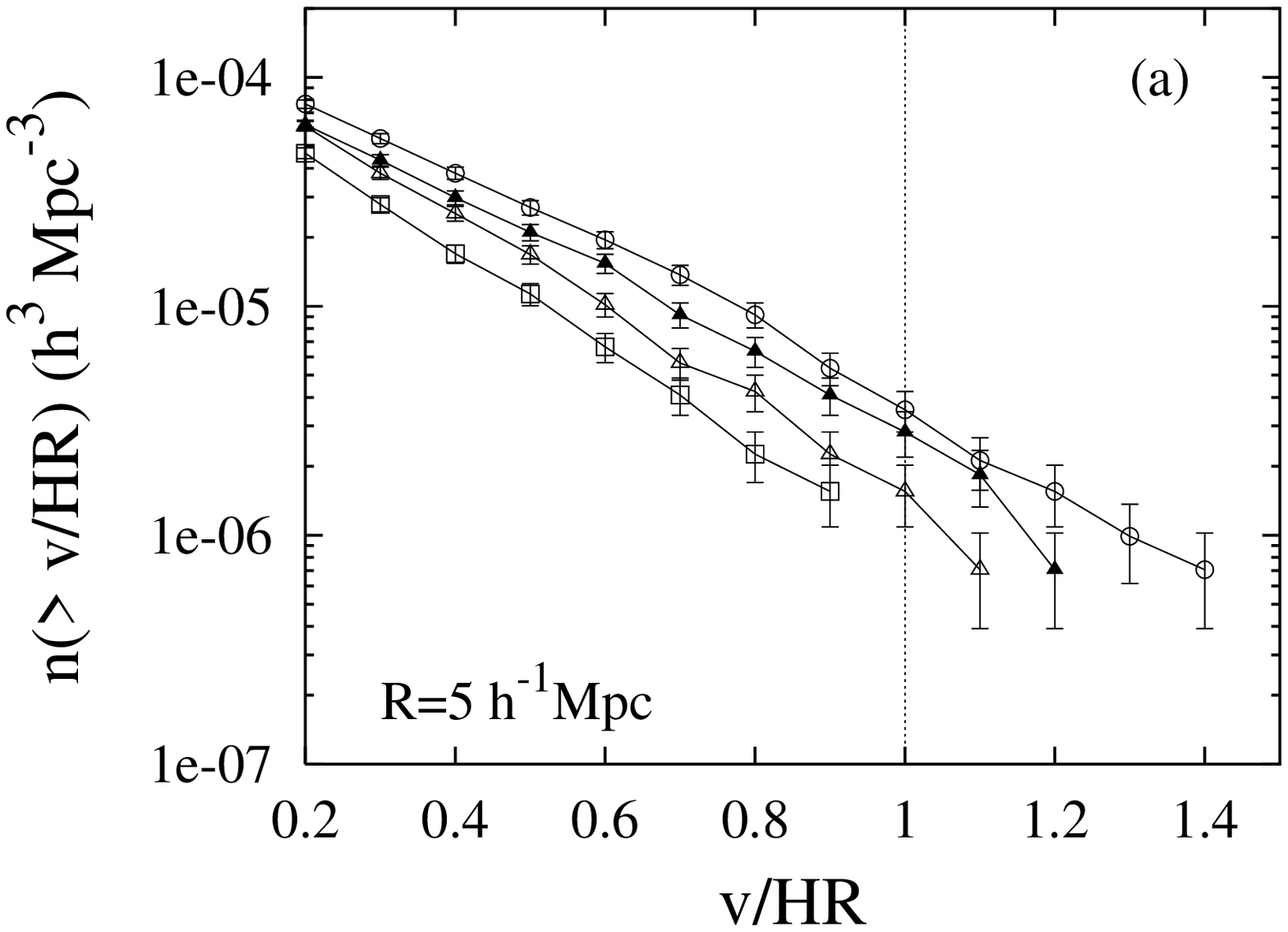,width=8cm}
\psfig{file=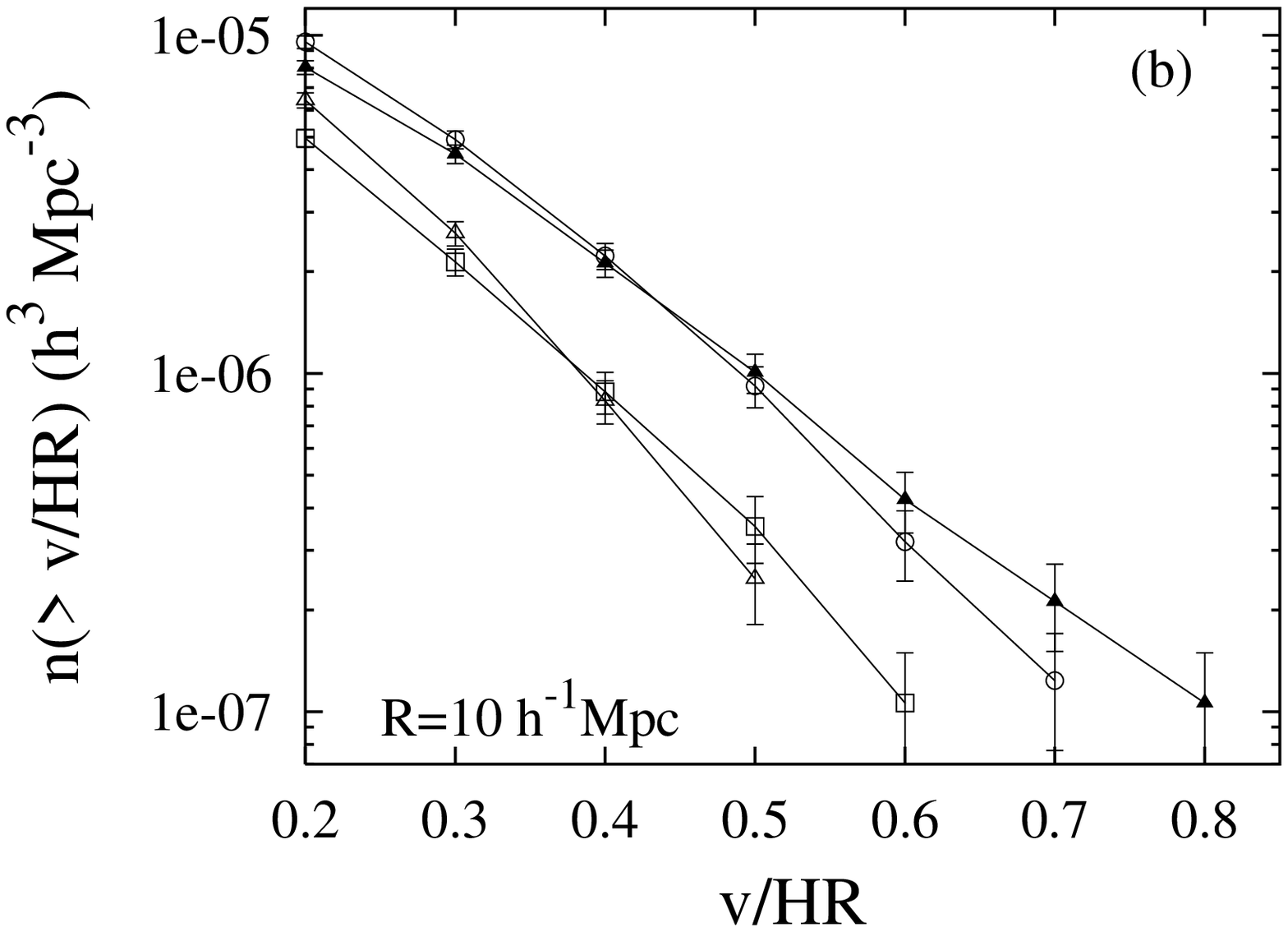,width=8cm}

\caption{(a) The velocity distribution for the superclusters in
the model S1 (squares), S2 (open triangles), S3 (filled triangles)
and S4 (circles). (b) The velocity distribution for the
superclusters in the model L1 (squares), L2 (open triangles), L3
(filled triangles) and L4 (circles). Error bars in panels a and b
denote Poisson errors.}

\end{figure}

Panel (b) in Fig.~8 shows the number density for the $10h^{-1}$Mpc
superclusters in the models L1-L4 ($L=384h^{-1}$Mpc). The velocity
distributions in the models L1 and L2 are similar. In the
model L1, the density parameter $\Omega_0$ is smaller, but the amplitude
of the large-scale density fluctuations is larger than in model L2 (see
Fig.~2). The number density of superclusters, which have peculiar
velocities $v>0.5 HR$, is $3.5 \times 10^{-7}h^3$Mpc$^{-3}$ and
$2.4 \times 10^{-7}h^{3}$Mpc$^{-3}$, in the models L1 and L2, respectively.
The velocity distributions in the models L3 and L4 are also similar. In
the model L4, the density parameter $\Omega_0$ is higher, but the
amplitude of the large-scale density fluctuations is smaller than in the
model L3. The number density of superclusters, which have peculiar
velocities $v>0.5 HR$, is $1.0 \times 10^{-6}h^3$Mpc$^{-3}$ and
$0.9 \times 10^{-6}h^{3}$Mpc$^{-3}$, in the models L3 and L4, respectively.

In the simulations studied, all the superclusters defined with
$10h^{-1}$Mpc smoothing are expanding by the present epoch. In the
models L3 ($\Omega_0=0.3$) and L4 ($\Omega_0=0.4$), the maximum
density contrast for the $R=10h^{-1}$Mpc superclusters is $11.2$
and $7.9$, respectively. The turnaround density contrast for the
$R=10h^{-1}$Mpc superclusters is expected to be similar or
somewhat smaller than for the $R=5h^{-1}$Mpc superclusters (see
Fig.~4 and Fig.~6). In the model M1 ($\Omega_0=0.3$) and S4
($\Omega_0=0.4$), the turnaround density contrast for the
$R=5h^{-1}$Mpc superclusters is $\delta_{ta}=16.2 \pm 4.6$ and
$\delta_{ta}=12.3 \pm 2.2$, respectively. Therefore, the
$R=10h^{-1}$Mpc superclusters in the models L3 and L4 are close
to the turnaround point.

\sec{SUMMARY} In this paper we have investigated the dynamical
state of superclusters in flat $\Lambda$CDM models. We
analyzed superclusters in the Virgo simulation for the
$\Lambda$CDM model with $\Omega_0=0.3$ and $\sigma_8=0.9$. This
simulation was carried out using the AP$^3$M code. We also
investigated superclusters in a series of N-body simulations
we carried out for flat $\Lambda$CDM models, where the density
parameter $\Omega_0=0.2-0.4$ and $\sigma_8=0.7-0.9$. In these
simulations, a PM code was used. 
To identify superclusters in simulations we used a method
where superclusters were defined as maxima of the density field
smoothed on the scale $R=10h^{-1}$Mpc. Smaller superclusters in
the density field smoothed on the scale $R=5h^{-1}$Mpc were also
examined. For each supercluster identified, we determined the
density contrast, $\delta$, and the radially averaged peculiar
velocity, $v$ (see eq. (23-24)).

The results for the superclusters in the N-body simulations were compared
with the spherical collapse model. We found that the radial peculiar
velocities in the N-body simulations are systematically smaller than
those predicted by the spherical collapse model ($\sim 15$\% and $\sim 25$\%
for the $R=10h^{-1}$Mpc and $R=5h^{-1}$Mpc superclusters, respectively).
The deviations from the spherical collapse model are similar for
$\delta \approx 1$ and $\delta \approx 10$. We also studied the
turnaround density contrast for superclusters. We found that
the turnaround density contrast for superclusters is
substantially larger than that predicted by the spherical collapse
model. In the Virgo simulation with $\Omega_0=0.3$, the
turnaround density contrast is $\delta_{\rm ta}=16.2\pm 4.6$ for
the $R=5h^{-1}$Mpc superclusters. For comparison, in the
spherical model the turnaround density contrast is $11.19$ for
$\Omega_0=0.3$.

We found the relations between $\delta$ and $v$ for different
cosmological models. These relations can be used to estimate the
dynamical state of a supercluster on the basis of its density
contrast. The density contrast for superclusters on large scales
can be determined from the new large galaxy redshift surveys.
The results of our study show how
the density contrast is related to the radial peculiar velocity
of the supercluster. For example, in the model with
$\Omega_0=0.3$, $\sigma_8=0.9$ and $R=10h^{-1}$Mpc, the peculiar
velocity $v/HR=0.54 \pm 0.04$ for the range $\delta=5.0-6.3$. The
$v/HR - \delta$ relation for different values of $R$ and
$\sigma_8$ is similar.

On the other hand, recent improvements of the techniques for
measuring distances of galaxies allow compilation of large
samples of galaxies with measured peculiar velocities. We can use
this information to study the radially averaged peculiar
velocity of superclusters. On the basis of this peculiar
velocity we can estimate the mass-density contrast in the
superclusters and compare it with the galaxy-density contrast. In
this way we can study the relation between the mass
distribution and the galaxy distribution on large scales.

In the simulations studied, all the superclusters defined with
$10h^{-1}$Mpc smoothing have peculiar velocities $v<HR$ and,
therefore, are expanding by the present epoch. A small fraction
of the $R=5h^{-1}$Mpc superclusters has already reached their
turnaround radius and these superclusters have started to
collapse. In the Virgo simulation, the number density of
superclusters, which have started to collapse, is $4.6 \times
10^{-6}h^3$Mpc$^{-3}$. The $R=10h^{-1}$Mpc superclusters in the
models with $\Omega_0=0.3$, $\sigma_8=0.9$ and $\Omega_0=0.4$,
$\sigma_8=0.8$ are close to the turnaround point.

\sec*{ACKNOWLEDGEMENTS}

We thank J. Einasto, M. Einasto, P. Heinam\"aki, G. H\"utsi, J. Jaaniste
and E. Saar for useful discussions. This work has been supported by
the ESF grant 3601. The Virgo simulation used in this paper is
available at {\it
http://www.mpa-garching.mpg.de/Virgo/virgoproject.html.} This
simulation was carried out at the Computer Center of the
Max-Planck Society in Garching and at the EPCC in Edinburgh, as
part of the Virgo Consortium project. The code CMBFAST used in
this paper, is available at {\it
http://arcturus.mit.edu/~matiasz/CMBFAST/cmbfast.html}.

\vfill

\begin{thebibliography}{}

\bibitem[\protect\citename{Bahcall}{1999}]{Bahcall99}
Bahcall N.A., Ostriker J.P., Perlmutter S., Steinhardt P.J., 1999,
Science, 1481

\bibitem[\protect\citename{Bardelli}{2000}]{Bardelli00}
Bardelli S., Zucca E., Zamorani G., Moscardini L., Scaramella R.,
2000, 312, 540

\bibitem[\protect\citename{Barmby}{1998}]{Barmby98}
Barmby P., Huchra J.P., 1998, AJ, 115, 6

\bibitem[\protect\citename{Basilakos}{2001}]{Basilakos01}
Basilakos S., Plionis M., Rowan-Robinson M., 2001,
MNRAS, 323, 47

\bibitem[\protect\citename{Batuski}{1999}]{Batuski99}
Batuski D.J., Miller C.J., Slinglend K.A., Balkowski C.,
Maurogordato S., Cayatte V., Felenbok P., Olowin R.,
1999, ApJ, 520, 491

\bibitem[\protect\citename{Bernardeau}{1994}]{Bernardeau94}
Bernardeau F., 1994, ApJ, 427, 51

\bibitem[\protect\citename{Bertschinger}{1991}]{Bertchinger91}
Bertschinger E., Gelb J.M., 1991, Comp. Phys., 5. 164

\bibitem[\protect\citename{Bond}{1984}]{Bond84}
Bond J.R., Efstathiou G., 1984, ApJ, 285, L45

\bibitem[\protect\citename{Bondi}{1947}]{Bondi47}
Bondi H., 1947, MNRAS, 107, 410

\bibitem[\protect\citename{Couchman}{1995}]{Couchman95}
Couchman H.M.P., Thomas P.A., Pearce F.R., 1995, ApJ, 452, 797

\bibitem[\protect\citename{Davis}{1982}]{Davis82}
Davis M., Huchra J., 1982, ApJ, 254, 437

\bibitem[\protect\citename{deVaucouleurs}{1956}]{deVaucouleurs56}
de Vaucouleurs G., 1956, Vistas Astron., 2, 1584

\bibitem[\protect\citename{Einasto}{1997}]{Einasto97}
Einasto M., Tago E., Jaaniste J., Einasto J., Andernach H., 1997,
A\&AS, 123, 119

\bibitem[\protect\citename{Eisenstein}{1995}]{Eisenstein95}
Eisenstein D.J., Loeb A., 1995, ApJ, 439, 520

\bibitem[\protect\citename{Eke}{1996}]{Eke96}
Eke V.R., Cole S., Frenk C.S., 1996, MNRAS, 282, 263

\bibitem[\protect\citename{Ettori}{1997}]{Ettori97} Ettori S.,
Fabian A.C., White D.A., 1997, MNRAS, 289, 787

\bibitem[\protect\citename{Freedman}{2001}]{Freedman01}
Freedman W.L. et al., 2001, ApJ, 553, 47

\bibitem[\protect\citename{Gramann}{1988}]{Gramann88}
Gramann M., 1988, MNRAS, 234, 569

\bibitem[\protect\citename{Gramann}{2001}]{Gramann01}
Gramann M., H\"utsi G., 2001, MNRAS, 327, 538

\bibitem[\protect\citename{Gunn}{1972}]{Gunn72}
Gunn J.E., Gott J.R., 1972, ApJ, 176, 1

\bibitem[\protect\citename{Hanski}{2001}]{Hanski01}
Hanski M.O., Theureau G., Ekholm T., Teerikorpi P., 2001,
A\&A, 378, 345

\bibitem[\protect\citename{Hockney}{1981}]{Hockney81}
Hockney R.W., Eastwood J.W. 1981, Numerical simulations using particles
(New York: McGraw-Hill)

\bibitem[\protect\citename{Jenkins}{1998}]{Jenkins98}
Jenkins A. et al. (The Virgo Consortium), 1998, ApJ, 499, 20

\bibitem[\protect\citename{Lahav}{1991}]{Lahav91}
Lahav O., Lilje P.B., Primack J.R., Rees M.J., 1991, 251, 128

\bibitem[\protect\citename{Lee}{1986}]{Lee86}
Lee H., Hoffman Y., Ftaclas C., 1986, ApJ, 304, L11

\bibitem[\protect\citename{Lilje}{1991}]{Lilje91}
Lilje P.B., Lahav O., 1991, ApJ, 374, 29

\bibitem[\protect\citename{Lokas}{2001}]{Lokas01}
Lokas E.L., Hoffmann Y., 2001, Proceedings of the 3rd International
Workshop on the Identification of Dark Matter, ed. N.J.C. Spooner
\& V. Kudryavtev, World Scientific, Singapore, p. 121

\bibitem[\protect\citename{OMeara}{2001}]{OMeara01}
O'Meara J.M., Tytler D., Kirkman D., Suzuki N., Prochaska J.X.,
Lubin D., Wolfe A.M., 2001, ApJ, 552, 718

\bibitem[\protect\citename{Ostriker}{1995}]{Ostriker95}
Ostriker J.P., Steinhardt P.J., 1995, Nat, 377, 600

\bibitem[\protect\citename{Parodi}{2000}]{Parodi00}
Parodi B.R., Saha A., Tammann G.A., Sandage A., 2000, ApJ, 540,
634

\bibitem[\protect\citename{Pearce}{1997}]{Pearce97}
Pearce F.R., Couchman H.M.P., 1997, NewA, 2, 411

\bibitem[\protect\citename{Peebles}{1980}]{Peebles80}
Peebles P.J.E., 1980, {\it The Large-Scale Structure of the Universe},
Princeton University Press

\bibitem[\protect\citename{Peebles}{1984}]{Peebles84}
Peebles P.J.E., 1984, ApJ, 284, 439

\bibitem[\protect\citename{Press}{1974}]{Press74}
Press W.H., Schechter P., 1974, ApJ, 187, 425

\bibitem[\protect\citename{Regos}{1989}]{Regos89}
Reg\"os E., Geller M.J., 1989, AJ, 98, 755

\bibitem[\protect\citename{Seljak}{1996}]{Seljak96}
Seljak U., Zaldarriaga M., 1996, ApJ, 469, 7

\bibitem[\protect\citename{Schechter}{1980}]{Schechter80}
Schechter P.L., 1980, AJ, 85, 801

\bibitem[\protect\citename{Shapley}{1930}]{Shapley30}
Shapley H., 1930, Harvard Obs. Bull., 874, 9

\bibitem[\protect\citename{Silk}{1974}]{Silk74}
Silk J., 1974, ApJ, 193, 525

\bibitem[\protect\citename{Silk}{1977}]{Silk77}
Silk J., 1977, A\&A, 59, 53

\bibitem[\protect\citename{Small}{1998}]{Small98}
Small T.A., Ma C., Sargent W.L.W., Hamilton D., 1998,
ApJ, 492, 45

\bibitem[\protect\citename{Suhhonenko}{1999}]{Suhhonenko99}
Suhhonenko I., Gramann M., 1999, MNRAS, 303, 77

\bibitem[\protect\citename{Suhhonenko}{2002}]{Suhhonenko02}
Suhhonenko I., Gramann M., 2002, MNRAS submitted, astro-ph:0203166

\bibitem[\protect\citename{Tolman}{1934}]{Tolman34}
Tolman R.C., 1934, Proc. Nat. Acad. Sci., 20, 169

\bibitem[\protect\citename{Tytler}{2000}]{Tytler00}
Tytler D., O'Meara J.M., Suzuki N., Lubin D., 2000,
Phys. Scr., 85, 12

\bibitem[\protect\citename{vanHaarlem}{1993}]{vanHaarlem93}
van Haarlem M., van de Weygaert R., 1993, ApJ, 418, 544

\bibitem[\protect\citename{Villumsen}{1986}]{Villumsen86}
Villumsen J.V., Davis M., 1986, ApJ, 308, 499

\bibitem[\protect\citename{Zeldovich}{1970}]{Zeldovich70}
Zel'dovich Ya. B., 1970, A\&A, 5, 84

\bibitem[\protect\citename{Yahil}{1985}]{Yahil85}
Yahil A., 1985, In {\it The Virgo Cluster}, ed. O. Richter \&
V. Bingelli, ESO, Garching, p. 359


\end{thebibliography}
\end{document}